\documentclass[12pt]{iopart}
\usepackage{graphicx}
\usepackage{amssymb}

\begin{document}

\title[]{Measurement of deep-subwavelength emitter separation in a waveguide-QED system}

\author{Zeyang Liao$^{1}$, M. Al-Amri$^{2}$,  and M. Suhail. Zubairy$^{1}$}

\address{$^{1}$Institute for Quantum Science and Engineering (IQSE) and Department of Physics and Astronomy, Texas A$\&$M University, College Station, TX 77843-4242, USA \\
$^{2}$The National Center for Applied Physics, KACST, P.O Box 6068, Riyadh 11442, Saudi Arabia}
\ead{zeyangliao@physics.tamu.edu}


\begin{abstract}
In the waveguide quantum electrodynamics (QED) system, emitter separation plays an important role for its functionality. Here, we present a method to measure the deep-subwavelength emitter separation in a waveguide-QED system. In this method, we can also determine the number of emitters within one diffraction-limited spot. In addition, we also show that ultrasmall emitter separation change can be detected in this system which may then be used as a waveguide-QED-based sensor to measure tiny local temperature/strain variation.    
\end{abstract}

%
\vspace{2pc}

\section{Introduction}

Photon transport in a waveguide system coupled to quantum emitters, known as ``waveguide-QED", has attracted extensive interests because of its possible applications in quantum device and quantum information \cite{Liao063004, Roy2017}. The static and dynamical solutions of the photon transport in the waveguide-QED system have been widely studied using various methods \cite{Shen2005a, Shen2005b, Yudson2008, Tsoi2008, Derouault2014, Kornovan2016,  Zheng063816, Roy2011a, Yan2011, Li063810, Fan063821, Lalumiere2013, Chen2013, Xu043845, Shi205111, Cheng2012, Pletyukhov095028, Shi053834, Chen2011, Liao2015, Zang2015, Roulet2016, Liao2016b}. Many possible applications have also been proposed such as atomic mirror/cavity \cite{Zhou2008, Chang2012, Guimond023808}, single-photon frequency comb generation \cite{Liao2016}, single-photon diode \cite{Menon2004, Shen173902}, single-photon transistor \cite{Chang2007, Witthaut2010, Tiecke2014, Kyriienko2016}, single-photon frequency converter \cite{Bradford103902, Bradford043814, Yan2013, Sun2014}, and quantum computation \cite{Ciccarello2012, Zheng2013b, Paulisch2016}. In addition, the waveguide-QED system is also a very good platform to study the many-body physics beacause long-range interaction is allowed in this system \cite{Douglas2015, Zheng2011, Fang2015}. Aside from the usual dielectric waveguide \cite{Dayan1062} and photonic crystal waveguide \cite{Englund2007}, these theories can be also applied to study the propagation of surface plasmon along the nanowire \cite{Akimov402} and the microwave photon along the superconducting transmission line \cite{ Wallraff2004, Abdumalikov193601, Hoi073601, Hoi263601, Loo1494}.

In the waveguide-QED system, collective interactions betweeen the emitters are critical for its functionality and they largely depend on the emitter separation \cite{Liao2015, Kornovan2016, Liao2016b, Li366, Cheng2017}. However, few has been discussed about how to measure the emitter separation in the waveguide-QED system especially when the emitters are in the deep-subwavelength scale. Besides, it is also an open question about how to determine the number of emitters coupled to the waveguide when the emitters are in a subwavelength region. Near-field scanning optical microscopy (NSOM) may be able to determine the number and positions of the emitters with high accuracy based on near-field point-by-point scanning  \cite{Betzig1993}. However, the near-field technique is surface bound and has limited applications. If the emitters are embedded inside a photonic crystal waveguide with low loss to the free space, NSOM may fail to detect the signal.

In this paper, we show a method to measure the emitter separation embedded in or side-coupled to a 1D waveguide. Our method can measure the emitter separation even if they are in the deep-subwavelength scale. We also show that the number of emitters within one diffraction-limited spot can be detmined from the number of reflection peaks. Our method shown here may also be used as super-resolution biosensing \cite{Chen39720}. In addition, since dipole-dipole energy shift is very sensitive to the emitter separation especially when they are in the deep subwavelength scale \cite{Ficek2004, Chang2006, Liao2012, Feng2017}, we may also detect ultrasmall emitter separation change from the emission spectrum shift. This may then be used to probe tiny temperature or strain change with high sensitivity. This waveguide-QED-based temperature/strain sensor can also have very high spatial resolution because the sensing region has a deep-subwavelength size. 

This paper is organized as follows: In Sec. II we show the schematic setup to measure the deep-subwavelength emitter separation and the emission spectra of the system. In Sec. III, we show how to measure the deep-subwavelength emitter separation in the cases when $\gamma=0$ and $\gamma\neq 0$. In Sec. IV, we show how to measure the ultrasmall emitter separation change in this system  which may then be used as a temperature sensor. In Sec. V, we show how to measure the coupling strength for non-identical emitter case. Finally we summarize the results.

\section{Model and Spectrum}

The schematic setup for measuring the deep-subwavelength emitter separation is shown in Fig. 1. Suppose that the two emitters have a spatial separation $d$. They can couple to each other via the guided and non-guided photon modes. The dipole-dipole coupling between the two emitters is distance-dependent and they can modify the scattering photon spectra of the waveguide-QED system. By monitoring the emitted photon spectrum, we can determine the emitter separation $d$ even if $d$ is much smaller than the resonant wavelength $\lambda$.

The interaction Hamiltonian of this system in the rotating wave approximation is given by \cite{Liao2016b}
\begin{equation}
H=\sum_{j=1}^{2}\hat{\sigma}_{j}^{+}[\hat{D}_g^{-}(\vec{r}_{j})e^{-i\Delta\omega_{k}^{j}t}+\hat{D}_{ng}^{-}(\vec{r}_{j})e^{-i\Delta\omega_{\vec{q},\lambda}^{j}t}]+H.c.,
\end{equation}
where $\hat{D}_{g}^{-}(\vec{r}_{j})=\sum_{k}\vec{\mu}_j\cdot \vec{E}^{g}_{k}(\vec{r}_{j})\hat{a}_{k}e^{ikz_j}$ and $D_{ng}^{-}(\vec{r}_{j})=\sum_{\vec{q},\alpha}\vec{\mu}_j\cdot \vec{E}^{ng}_{\vec{q},\alpha}(\vec{r}_{j})\hat{a}_{k}e^{i\vec{q}\cdot \vec{r}_{j}}$ are the coupling coefficients with the guided and non-guided modes, respectively. Here, $\sigma_{j}^{+}$ is the raising operator of the $j$th emitter with position $\vec{r}_{j}$ ($z_{j}$ is its $z$ component along the waveguide direction), and $\vec{\mu}_j$ is the transition dipole moment. $\vec{E}^{g}_{k}(\vec{r}_{j})$ is the electric field strength of the waveguide mode with wavevector $k$ at position $\vec{r}_{j}$, and $\vec{E}^{ng}_{\vec{q},\alpha}(\vec{r}_{j})$ is the electric field of the non-guided mode with wavevector $\vec{q}$ and polarization $\alpha$ at position $\vec{r}_{j}$. $\Delta\omega_{k}^{j}=\omega_{k}-\omega_{j}$ ($\Delta\omega_{\vec{q},\alpha}^{j}=\omega_{\vec{q}, \alpha}-\omega_{j}$) is the detuning between the transition frequency of the jth emitter $\omega_{j}$ and the frequency of the guided photon $\omega_{k}$ (non-guided photon modes $\omega_{\vec{q}, \alpha}$).

\begin{figure}
\includegraphics[width=0.98\columnwidth]{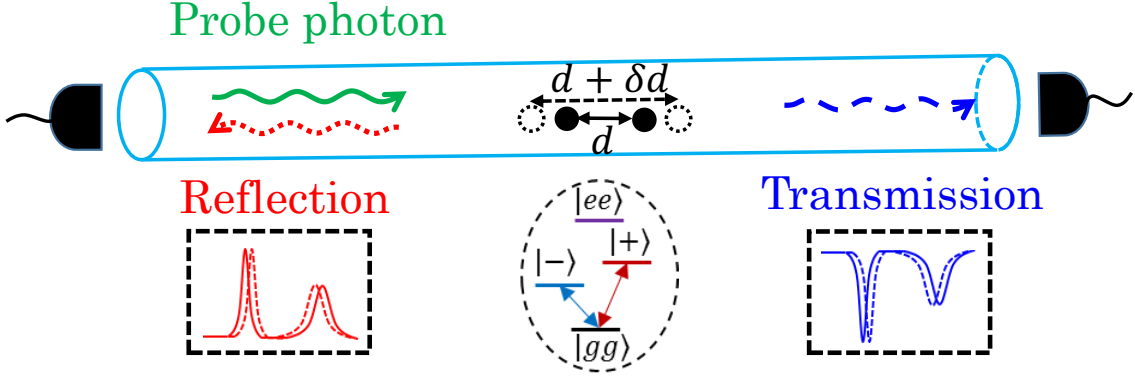}
\caption{ Measurement of deep-subwavelength emitter separation and separation change in a waveguide-QED system.  } 
\end{figure}

When we send a single photon pulse into this system, the reflection and transmission amplitudes are given by \cite{Liao2016b}
\begin{eqnarray}
r(\delta k)&=&e^{2ikz_{1}}\frac{2M_{12}(\delta k)e^{ikd}-M_{11}(\delta k)e^{2ikd}-M_{22}(\delta k)}{M_{11}(\delta k)M_{22}(\delta k)-M_{12}^{2}(\delta k)}, \\
t(\delta k)&=&1-\frac{M_{11}(\delta k)+M_{22}(\delta k)-2M_{12}(\delta k)\cos(kd)}{M_{11}(\delta k)M_{22}(\delta k)-M_{12}^{2}(\delta k)},
\end{eqnarray}
where $k=k_a+\delta k$ with $k_a=2\pi/\lambda$, $M_{ii}(\delta k)=(1+\gamma_{i}/\Gamma_{i})+2i(\Delta \omega_{i}-\delta k v_{g})/\Gamma_{i}$ and $M_{12}(\delta k)=2V_{12}e^{ikd}/\sqrt{\Gamma_{1}\Gamma_{2}}$. Here, $\Gamma_{i} (\gamma_{i})$ is the spontaneous decay rate of the ith emitter due to the guided (non-guided) photon modes. $V_{12}e^{ik_{a}d}$ is the dipole-dipole interaction strength between the two emitters with $V_{12}$ given by
\begin{equation} 
V_{12}=\frac{\sqrt{\Gamma_{1}\Gamma_{2}}}{2}+\frac{3\sqrt{\gamma_{1}\gamma_{2}}}{4}\Big [\frac{-i}{k_{a} d}+\frac{1}{(k_{a} d)^2}+\frac{i}{(k_{a} d)^3}\Big ].
\end{equation}
Here, we assume that the incident photon has a transverse polarization (TE-mode) so that the induced transition dipole moment is perpendicular to the waveguide direction. From Eq. (4), we see that the dipole-dipole interaction is highly distance-dependent. In the following, we first show how to determine the deep-subwavelength emitter separation from the spectrum characteristics in the cases with or without non-guided photon modes.  Then we show how to measure ultrasmall emitter separation change from the emission spectrum shift. Since we mainly consider the case when the emitters have deep-subwavelength distance, we can safely assume that the emitters feel the same local field and have the same decay rates, i.e, $\Gamma_{1}=\Gamma_{2}\equiv \Gamma$ and $\gamma_{1}=\gamma_{2}\equiv \gamma$. In Sec. V, we consider the case when $\Gamma_{1}\neq \Gamma_{2}$ and $\gamma_{1}\neq \gamma_{2}$.

\begin{figure}
\includegraphics[width=\columnwidth]{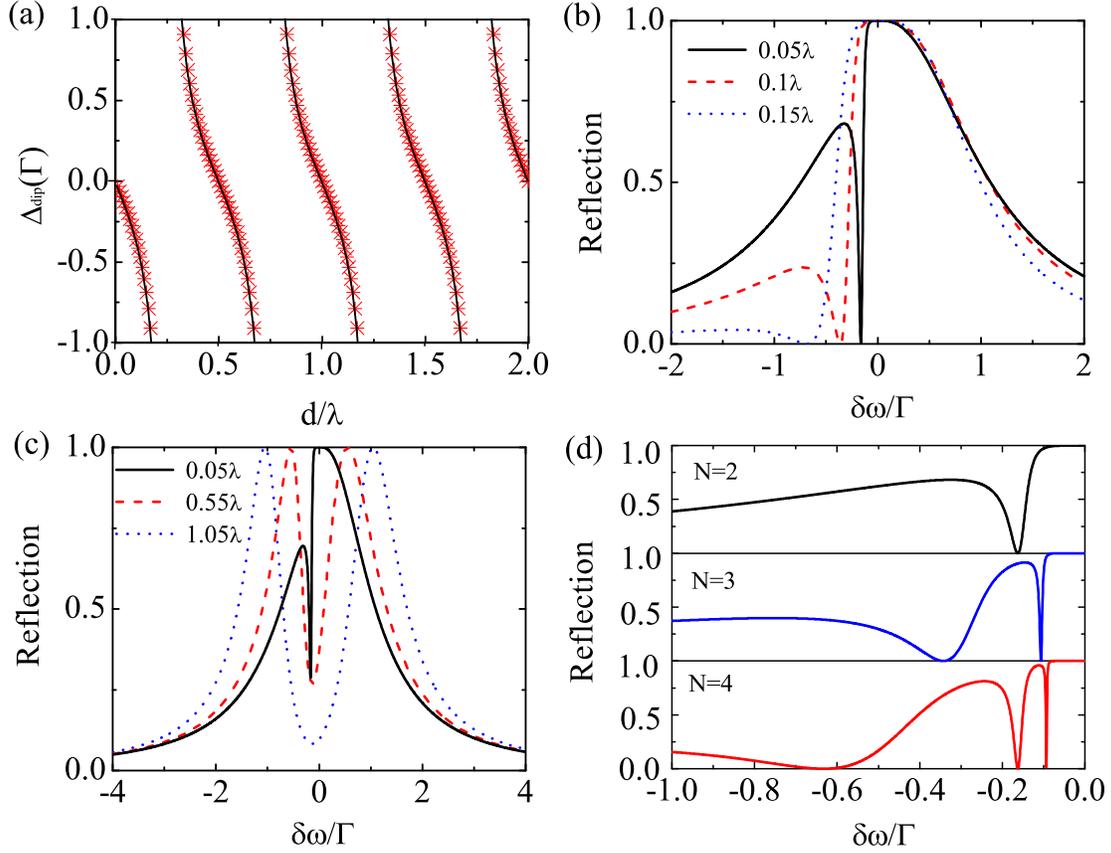}
\caption{(a) The reflection dip frequency as a function of emitter separation. The red asterisk symbol is the numerical calculated dip positions and the black solid line is the fitting of $-\tan(k_{a}d)/2$. (b) The reflection spectra for three different subwavelength emitter separation ($d=0.05\lambda, 0.1\lambda, 0.15 \lambda$). (c) The reflection spectra for three emitter separations ($d=0.05\lambda, 0.55\lambda, 1.05\lambda$) when a gradient electric field is applied. (d) The reflection spectra when there are 2, 3, and 4 emitters within one diffraction-limited spot. The emitter separation $d=0.05\lambda$. }
\end{figure}

\section{Measuring deep-subwavelength emitter separation}

\subsection{$\gamma=0$}

We first consider a high quality 1D waveguide where non-guided modes are negligible, i.e., $\gamma=0$. In this case, $M_{11}(\delta k)=M_{22}(\delta k)=1-2i\delta k v_{g}/\Gamma$ and $M_{12}=e^{ikd}$. The reflection amplitude is then given by
\begin{equation}
r(\delta k)=e^{2ikz_{1}}\frac{2e^{2ikd}-(1-2i\delta k v_{g}/\Gamma)(1+e^{2ikd})}{(2i\delta k v_{g}/\Gamma)^{2}-e^{2ikd}},
\end{equation}
from which it is not difficult to see that the reflection vanishes under the condition
\begin{equation}
\delta\omega\equiv\delta k v_{g}=-\frac{\Gamma}{2}\tan(k_{a}+\delta k)d.
\end{equation}
Since $\delta k$ is usually much less than $k_{a}$, the reflection vanishes at the frequency approximately equals to $-\Gamma \tan(k_{a}d)/2$ which is a function of emitter separation. The reflection dip frequency as a function of emitter separation is shown in Fig. 2(a) where the red asterisk symbol is the exact numerically calculated result from Eq. (6) and the black solid line is the fitting of $-\Gamma\tan(k_{a}d)/2$. We can see that the theoretical prediction matches the result calculated by  Eq. (6) very well. Hence, from the dip position $\Delta_{dip}$ we can determine the emitter separation $d=n\lambda /2+\lambda \textrm{arctan}^{-1}(2\Delta_{dip}/\Gamma)/2\pi$ where $n=0,1,\cdots$. The existence of $n$ is because the dipole-dipole coupling induced by the guided modes is a periodical function. Additional procedure is required to fix $n$. For example, a gradient field can be applied to distinguish the results for different $n$.

The reflection spectra for three different emitter separations ($d=0.05\lambda, 0.1\lambda, 0.15 \lambda$) are shown in Fig. 2(b). Indeed, there is a reflection dip in all three reflection spectra and the dip positions are different for different emitter separations. When the emitter separation increases from $d=0.05\lambda$ to $d=0.15\lambda$, the dip frequency is shifted from $-0.162\Gamma$ to $-0.688\Gamma$. 
From the dip frequency $-0.162\Gamma$ we can determine that the emitter separation is $0.05\lambda$, and from  $-0.688\Gamma$  we can determine that the emitter separation is $0.15\lambda$. Both results match the given value very well. However, we should mention that the solution is not unique because $n\lambda/2+d$ also gives the same spectrum. This problem can be solved by applying a gradient electric or magnetic field to shift the transition frequency of the emitters. For a fixed gradient field, different emitter separation has different energy shift. For example, the reflection spectra when $d=(n/2+0.05)\lambda$ with $n=0, 1, 2$ are shown in Fig. 2(c) where the energy shift due to the gradient field is $2.0\Gamma/\lambda$. When $d=0.05\lambda$, the energy difference between the two emitters is $0.1\Gamma$ which is relatively small. The reflection spectrum is shown as the red solid curve in Fig. 2(c) where we see that the spectrum is similar to the case when there is no gradient field. However, when $d=0.55\lambda$, the energy difference between the two emitters is $1.1\Gamma$ and the reflection spectrum is quite different from the case without gradient field. There are two reflection peaks with similar shape and the separation between the two peaks is measured to be $1.1\Gamma$. From this separation, we can determine  the emitter separation $d=0.55\lambda$. Similarly, when $d=1.05\lambda$, there are also two reflection peaks with separation $2.1\Gamma$ from which we can determine that $d=1.05\lambda$.

When there are more than two emitters which are very close to each other, the spectrum can become more complicated and a full spectrum fitting may be required to extract all the emitter separations. Nonetheless, the number of emitters can be determined by simply counting the number of dips in the reflection spectra.  The examples when there are 2, 3, and 4 emitters are shown in Fig. 2(d) where we can see that there is one dip for 2 emitters, two dips for 3 emitters, and three dips for 4 emitters. Thus the number of emitters is equal to the number of dips in the reflection spectra plus one.

\subsection{$\gamma\neq 0$}

We then consider the case when the non-guided modes are not negligible, i.e., $\gamma \neq 0$. This can be applied to the case when the waveguide is lossy or the emitters are side-coupled to the waveguide. In this case, the reflection amplitude shown in Eq. (2) can be rewritten as
\begin{equation}
r(\delta k)=\frac{A(\delta k)}{M_{11}(\delta k)+M_{12}(\delta k)}+\frac{B(\delta k)}{M_{11}(\delta k)-M_{12}(\delta k)}
\end{equation}
where $A(\delta k)=-e^{2ikz_{1}}(1+e^{ikd})^{2}/2$ and $B(\delta k)=-e^{2ikz_{1}}(1-e^{ikd})^{2}/2$. For deep-subwavelength ($d\ll \lambda$) region, $\delta k d\simeq 0$ and $kd\simeq k_a d$. We have $A(\delta k)\simeq -e^{2ikz_{1}}(1+e^{ik_a d})^{2}/2$, $B(\delta k)\simeq -e^{2ikz_{1}}(1-e^{ik_a d})^{2}/2$, and $M_{12}(\delta k)\simeq 2V_{12}e^{ik_{a}d}/\Gamma$. The denominators in Eq. (7) can be simplified as
\begin{equation}
M_{11}(\delta k)\pm M_{12}(\delta k)= (2/\Gamma)\{-i[\delta kv_{g} -\omega_{\pm}]+\Gamma_{\pm}\}
\end{equation}
where $\omega_{\pm}\simeq \pm\textrm{Im}[V_{12}e^{ik_{a}d}]$ is the collective frequency shift due to the dipole-dipole interaction and $\Gamma_{\pm}\simeq(\Gamma+\gamma)/2 \pm \textrm{Re}[V_{12}e^{ik_{a}d}]$ is the collective decay rate. Thus, there are two reflection peaks at the frequencies $\omega_{\pm}$ with the full width at half maximum (FWHM) linewidth $2\Gamma_{\pm}$. In the deep-subwavelength region, the reflectivity at the superradiant peak $\omega_{+}$ is about $\Gamma^{2}/\Gamma_{+}^{2}$ with $\Gamma_{+}\simeq \Gamma+\gamma$.  From the reflection peak positions, the linewidths and the reflectivity at the superradiant peak, we can obtain the dipole-dipole energy shifts and the collective decay rates.  Since both values depend on the emitter separation, it becomes possible that their separation can be extracted from the spectrum. 

Typical reflection spectrum is shown in Fig. 3(a) where we see that there are two reflection peaks with one superradiant peak and the other one subradiant peak. From the maximum reflectivity at the superradiant peak we can determine the ratio between $\Gamma$ and $\gamma$. Together with the condition $\Gamma_{+}+\Gamma_{-}=\Gamma+\gamma$, we can then determine $\Gamma$ and $\gamma$ separately. The difference of the two reflection linewidths gives $\Gamma_{+}-\Gamma_{-}=-2 \textrm{Re}[V_{12}e^{ik_{a}d}]$. The reflection peak position $\omega_{p}=\textrm{Im}[V_{12}e^{ik_{a}d}]$. From these two equations, we can determine the emitter separation by searching a separation $d$ which minimizes the variance $\delta S=\{[\textrm{Re}(V_{12}e^{ik_{a}d})-(\Gamma_{+}-\Gamma_{-})/2]^2+[\textrm{Im}(V_{12}e^{ik_{a}d})-\omega_{p}]^2\}/\gamma_{0}^{2}$.

In Fig. 3(a), we show the reflection spectra for two different emitter separations ($d=0.05\lambda$ and $d=0.08\lambda$). For both spectra, the reflectivity at the superradiant peak is $0.44$ from which we can obtain $\Gamma^{2}/(\Gamma+\gamma)^2=0.44$.  Thus, we can obtain $\Gamma=1.97\gamma$ which is very close to the given value $2\gamma$.  The spectrum when $d=0.05\lambda$ is shown as the red solid line. From the spectrum, we can measure the two peak positions to be $\pm 23.388\gamma_{0}$ and their linewidths are $5.88\gamma_{0}$ and $0.12\gamma_{0}$, respectively. From the summation of the two linewidths, we have $\Gamma+\gamma=3\gamma_{0}$. Since $\Gamma=1.97\gamma$, we can obtain $\Gamma=1.99\gamma_{0}$ and $\gamma=1.01\gamma_{0}$. From the difference of the two linewidths, we have $\textrm{Re}[V_{12}e^{ik_{a}d}]=1.44\gamma_{0}$. From the peak position we have $\textrm{Im}[V_{12}e^{ik_{a}d}]=23.388\gamma_{0}$. By searching the parameters $d$ such that $\delta S=(\textrm{Re}[V_{12}e^{ik_{a}d}]-1.44\gamma_{0})^2+(\textrm{Im}[V_{12}e^{ik_{a}d}]-23.388\gamma_{0})^2/\gamma_{0}^2 $ is minimized, we obtain that $d=0.0502\lambda$ with variance $\delta S=0.003$. We can see that the emitter separation is very close to the actual values $0.05\lambda $ with an error $0.4\%$.

The reflection spectra when $d=0.08\lambda$ is shown as the blue dashed line in Fig. 3(a). The superradiant peak is at frequency $5.732\gamma_{0}$ with the linewidth $5.76\gamma_{0}$. The subradiant peak is at frequency $-5.752\gamma_{0}$ with the linewidth $0.32\gamma$. From the summation of the two linewidths we have $\Gamma+\gamma=3.04\gamma_{0}$. We can then determine that $\Gamma=2.02\gamma_0$ and $\gamma=1.02\gamma_0$ which is also very close to the given values. From the difference of the two linewidths, we have $\textrm{Re}[V_{12}e^{ik_{a}d}]=1.36\gamma_0$. The superadiant peak position is slightly different from the subradiant peak position. We can use their average to estimate the imaginary part of the dipole-dipole interaction, i.e., $\textrm{Im}[V_{12}e^{ik_{a}d}]=5.742\gamma_0$. By least square fitting, we can determine that $d=0.0808\lambda$ with $\delta S=0.0003$. The extracted emitter separation is very close to the real separation $0.08\lambda $ with an error being about $1\%$.  Hence, the deep-subwavelength emitter separation can be extracted with very high accuracy.

\begin{figure}
\includegraphics[width=0.49\columnwidth]{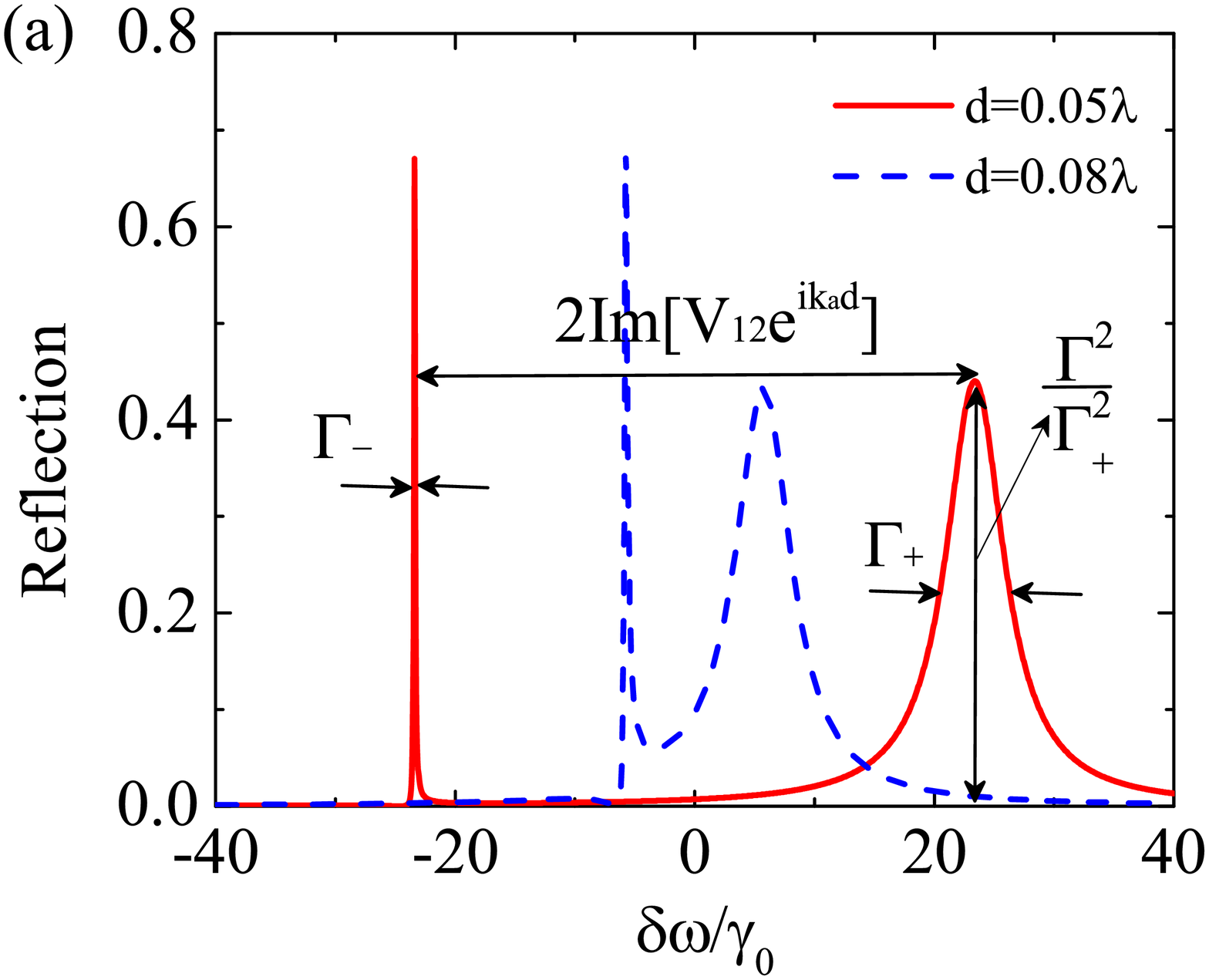}
\includegraphics[width=0.49\columnwidth]{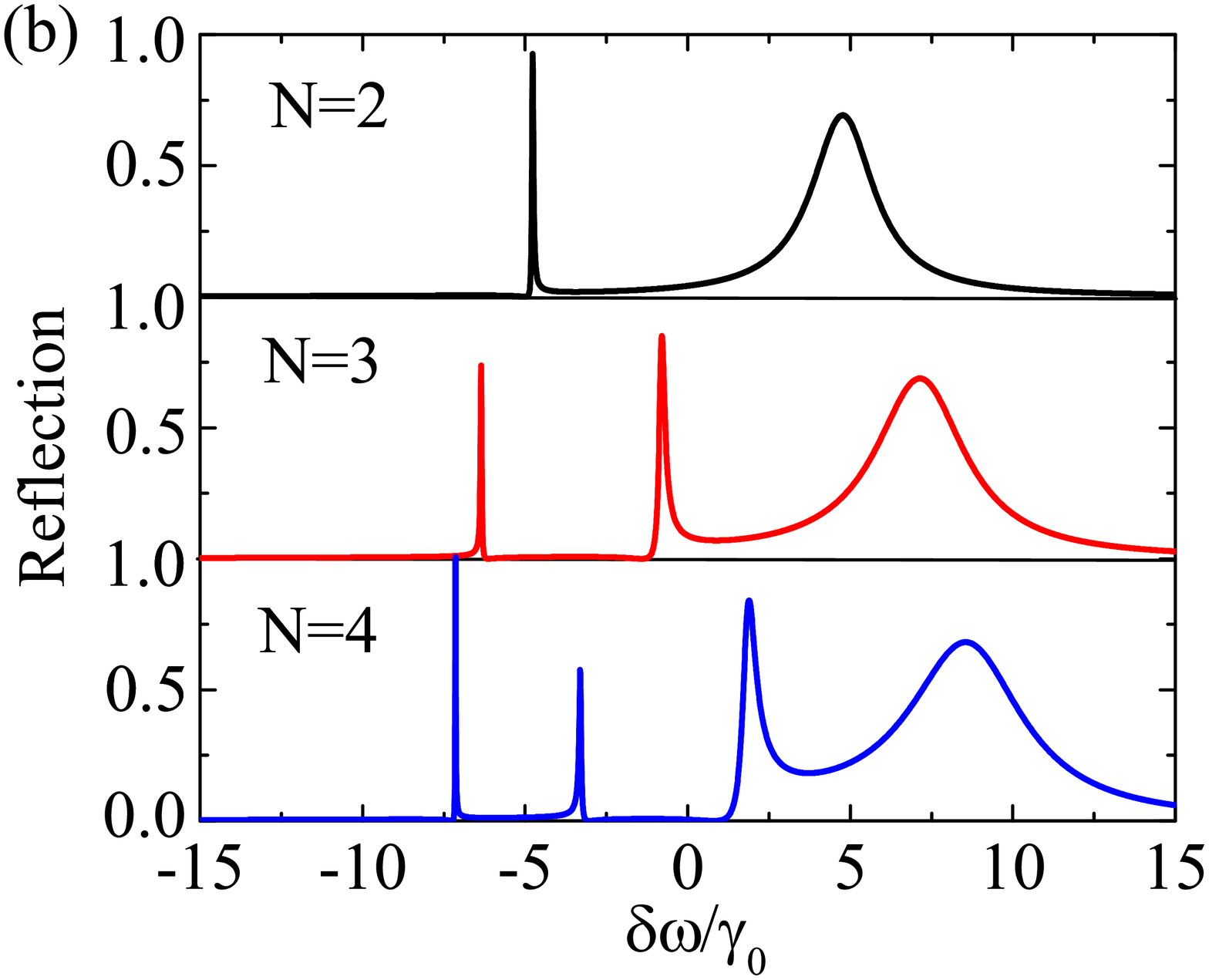}
\caption{(a) The reflection spectra for two different sub-wavelength emitter separations ($d=0.05\lambda$: red solid line; $d=0.08\lambda$: blue dashed line) when $\Gamma=2\gamma=2\gamma_0$.  (b) The reflection spectra when there are 2, 3, and 4 emitters with neighboring emitter separation being $0.05\lambda$ and $\Gamma=5\gamma=5\gamma_0$. }
\end{figure}

Similar to the case of perfect waveguide, we can also determine the number of emitters within one diffraction-limited spot by simply counting the number of reflection peaks in the reflection spectra. The numerical examples for 2, 3, and 4 emitters are shown in Fig. 3(c) where we can see that the number of reflection peaks equals to the number of emitters.

\section{Measuring emitter separation change and temperature sensing}

In the previous section, we show that deep sub-wavelength emitter separation in a waveguide-QED system can be measured from the emission spectrum. Here we show that we can also measure ultrasmall emitter separation change in a waveguide-QED system from the reflection spectrum shift.  


\begin{figure}
\includegraphics[width=0.48\columnwidth]{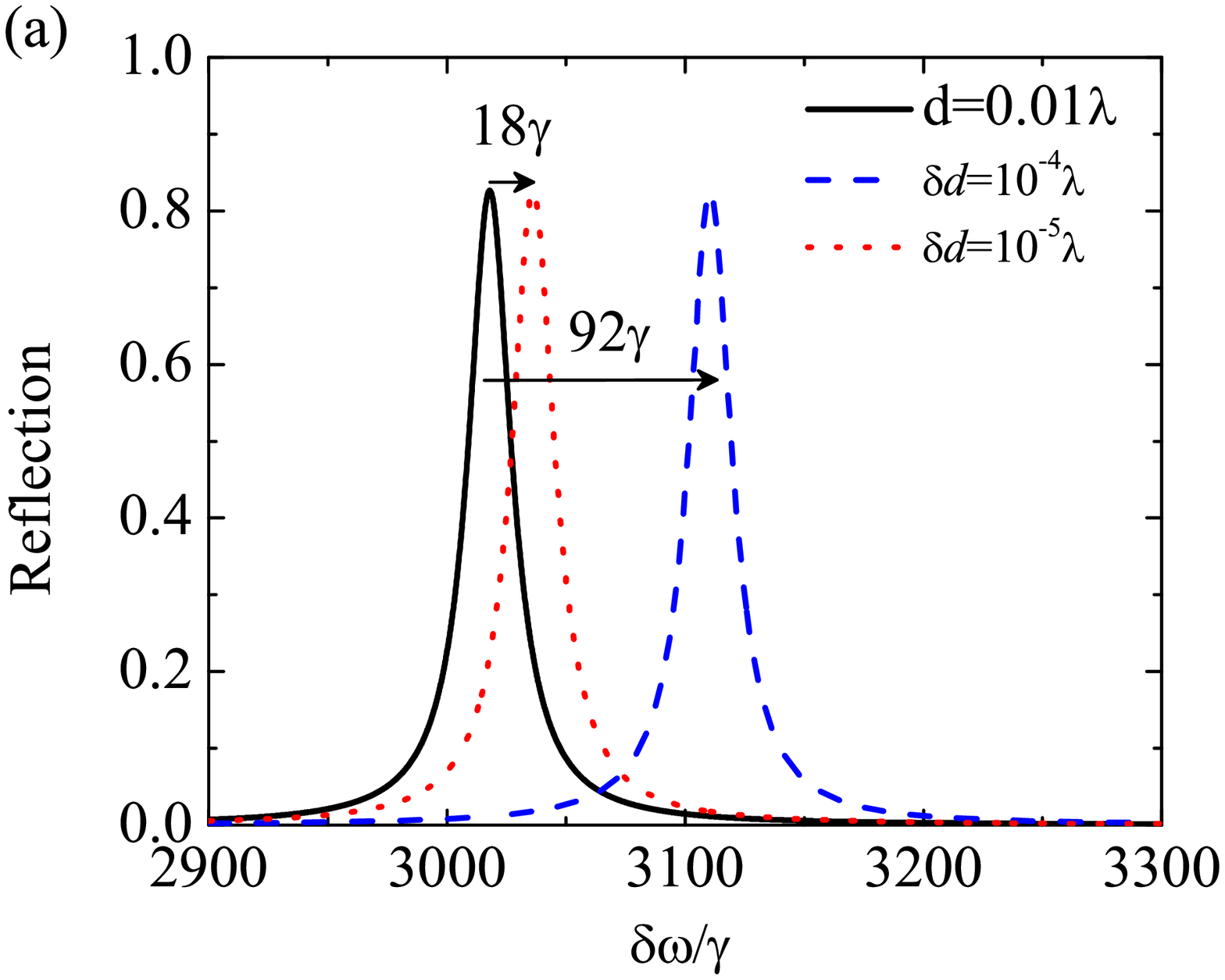}
\includegraphics[width=0.48\columnwidth]{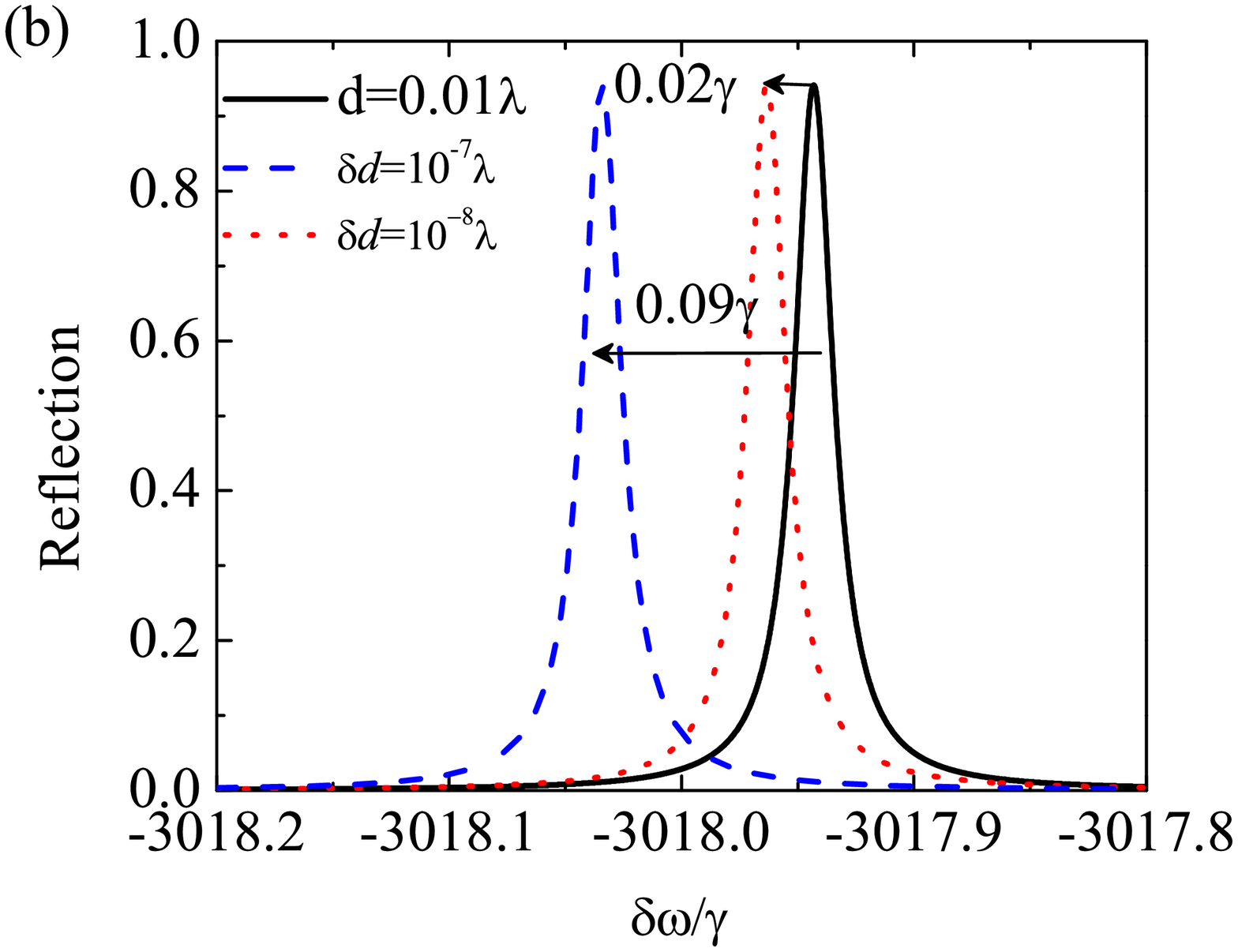}
\caption{Measurement of ultrasmall emitter separation change from spectrum shift.  (a) Superradiant peak shift. (b) Subradiant peak shift. The solid lines are the reflection spectra when $d=0.01\lambda$. The red dotted and blue dashed lines are the reflection spectra for the separation changes shown in the figure. Here $\Gamma=10\gamma$. }
\end{figure}

From Eq. (4) we see that the dipole-dipole energy shift $\Omega_{12}\approx 3\gamma/4(k_{a}d)^3$ when $d\ll \lambda$. The frequency shift caused by emitter separation change is given by $\delta\omega \approx -3\Omega_{12}\delta d/d$. When $d$ is small, $\Omega_{12}$ can be very large and $\delta\omega$ can be very large for even very small $\delta d$.  In the fiber Bragg grating sensor, the frequency shift due to the effective waveguide expansion is given by $\delta \omega_{FBG} \approx -\omega_{a}\delta d/\lambda$ \cite{Bock2006}. The ratio of the sensitivity between our method and the FBG sensor is given by $\delta\omega /\delta \omega_{FBG}\sim 3\lambda \Omega_{12}/(\omega_{a}d)$. To increase the sensitivity of our method, we need to reduce the emitter separation $d$ and increase $\Omega_{12}$ as large as possible. Supposing that $\gamma=10^{-7}\omega_{a}$ and $d=3\times 10^{-3}\lambda$, we have $\Omega_{12}\approx 10^{-2}\omega_{a}$ and the sensitivity is enhanced by about 10 times. For $\lambda\sim 1\mu$m, $d\sim 3$nm, so the sensing region has a nanometer size. Hence, our method shown here can have both high sensitivity and spatial resolution. Because the strain or temperature can modify the effective reflection index or cause thermal expansion of the waveguide, measurement of the effective emitter separation change may also be used to detect the tiny strain or temperature variation.

The refection spectrum shifts due to the emitter separation changes when $d\simeq 0.01\lambda$ are shown in Fig. 4 where Fig. 4(a) is the superradiant reflection spectrum shift and Fig. 4(b) is the subradiant relfection spectrum shift. The black solid curves in Fig. 4(a) and 4(b) are the results when 
$d=0.01\lambda$. When the emitter separation is reduced by an amount $\delta d=10^{-4}\lambda$, the center of the superradiant peak is shifted by $92\gamma$ (blue dashed line in Fig. 4(a)). From this frequency shift, we can determine that the effetive emitter separation change $\delta d=\delta \omega d/2\Omega_{12}\simeq 10^{-4}\lambda$. The effective length change caused by strain in a silica fiber for $\lambda=1.55\mu$m is about $1.25pm/\mu\varepsilon$, while it is $12.5pm/K$ for temperature change  \cite{Bock2006}.   For $\lambda=1.550\mu$m, $\delta d= 155pm$. which corresponds to a change $124$ microstrain or $12.4$K temperature. The minimum distance change we can measure from the superradiant peak shift is about $2\times 10^{-5}\lambda$ (red dotted line in Fig. 4(a)) where the shift of the superradiant peak is about $18\gamma$ and the reflection spectrum has significant overlap with the original spectrum. In this case, it corresponds to a change of 25 microstrain or $2.5$K temperature. However, if we probe the subradiant peak, the sensitivity can be much higher because the subradiant peak has much narrower linewidth. As shown in Fig. 4(b), when the emitter separation change $\delta d\sim 10^{-7}\lambda$, the subradiant peak shift is about $0.09\gamma$.  Despite that this frequency shift is small, it is well separated from the original reflection peak and it can be probed by a laser with narrow linewidth. The minimum frequency shift we can distinguish is about $0.02\gamma$ which correspons to a separation change of $2\times 10^{-8}\lambda$. When $\lambda=1.55\mu$m, it corresponds to a change of $0.03$pm which corresponds to a change of $0.02$ microstrain or $2$mK temperature. In this example, if $\gamma\sim 100$MHz such as in silicon vacancy center \cite{Rogers2014}, the minimum frequency shift is about $2$MHz. In practice, we can use a continuously tunable laser with narrow linewidth to probe this frequency shift \cite{Ball1992}.  However, in practice the atom in a crystal can have thermal vibration at finite temperature. The vibration amplitude at room temperature is of the order of $10^{-11}\sim 10^{-12}$m \cite{Lonsdale1948}. Due to this vibration, the smallest temperature variance we can measure is limited to be about $0.1^{o}$C. At very low temperature, the phonons are mostly long-wavelength acoustic phonon where the nearby lattice site can have almost the same vibration direction and amplitude. In this case, the emitter separation can remain almost the same despite that they can vibrate a bit. Since the dipole-dipole interaction depends on their relative separation instead of their absolute positions, the effect of the thermal vibrations to the sensor can be reduced and the sensitivity can be increased further.

Our waveguide-QED-based sensor may find important applications in various areas because temperature plays a crucial role in many areas such as material formation and biochemical reaction \cite{Lowell2000}. There are a number of techniques for measuring temperature. The conventional thermometers, such as thermography and thermocouples, suffer from a lack of sensitivity and spatial resolution. Fluorescent polymeric thermometer (FPT), which is based on the temperature-dependent fluorescence life time, can achieve high sensitivity \cite{Ookabe2012}. However, FPT is working on the diffraction-limited scale and the spatial resolution is not very high. The fiber Bragg grating (FBG) sensor can also measure ultrasmall temperature change, but its spatial resolution is also diffraction-limited \cite{Rao1997, Kersey1997, Lee2003, Bock2006}. In comparison, our proposed waveguide-QED-based temperature sensor has sensitivity the same as the FBG, but with much higher spatial resolution where the sensing region can be in the nanometer region. The temperature sensing based on nitrogen-vacancy (NV) center can have both high sensitivity (tens of milliKevin) and high spatial resolution (nanometer region) \cite{Neumann2013, Kucsko2013, Clevenson2015, Fedotov2015}. However, one disadvantage of NV-center-based temperature sensing is that the microwave and laser applied may significantly alter the original sample temperature. Although the sensitivity of our waveguide-QED-based temperature sensor may have less sensitivity than that based on NV-center, our method has its own advantage.  The probe photon in our temperature sensor is at the single-photon level and it is mostly confined inside the waveguide. Only a small portion of the evanescent light can couple to the sensing region, so the modification of the sample temperature by the probe light is minimized.

\section{Non-identical coupling strength} 

\begin{figure}
\centering
\includegraphics[width=0.8\columnwidth]{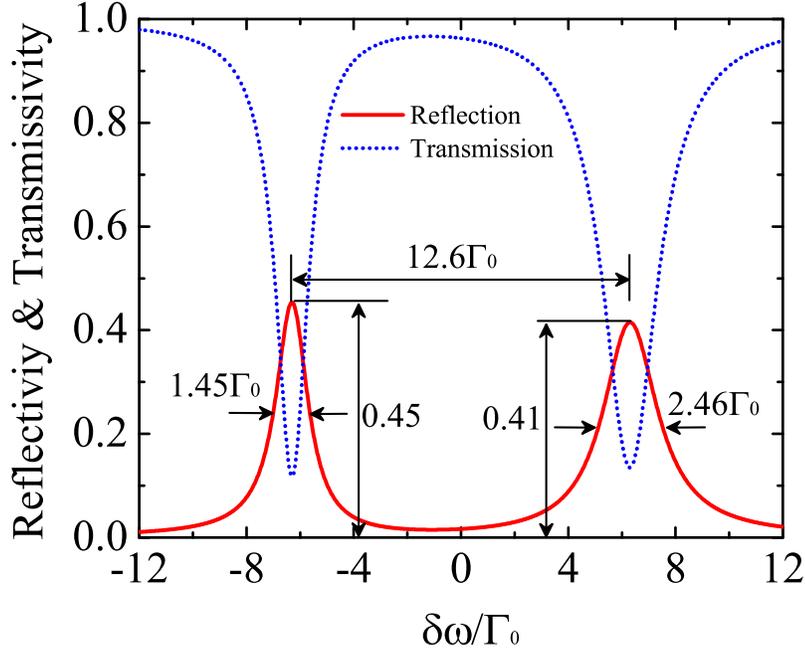}
\caption{The reflection and transmission spectra of two-emitter system under gradient field. The emitter separation is $2.1\Gamma_0$ and the coupling constants are given by $\Gamma_1=\Gamma_0, \gamma_1=0.5\Gamma_0, \Gamma_2=1.5\Gamma$, and $\gamma_2=0.9\Gamma_0$. }
\end{figure}

In the previous sections, we assumed that the emitters have the same coupling strength with both the waveguide and non-waveguide modes. This assumption is valid when the two emitters are spatially very close to each other, i.e., they can feel the same local field. However, when the two emitters are not in the sub-wavelength distance, the two emitters can have different coupling strengths with the guided and non-guided fields. In the section, we show a method to measure the coupling strengths and decay rates of the two emitters when they are different. 

We have shown that the reflectivity of a single emitter at the resonant frequency is given by $(1+\gamma/\Gamma)^{-2}$, and the linewidth of reflection spectrum is given by $\Gamma+\gamma$ \cite{Liao2015}. From these two characteristic parameters, we can determine $\Gamma$ and $\gamma$, respectively. For the two-emitter case, we can first apply a gradient electric or magnetic field to distinguish the two emitters. If the energy difference between the two emitters is greater than their coupling strength, they can be treated as separate emitters and then we can determine the two coupling strengths separately. For example, we assume that the distance between the two emitters is $2.1\lambda$ with $\Gamma_1=\Gamma_0, \gamma_1=0.5\Gamma_0, \Gamma_2=1.5\Gamma$, and $\gamma_2=0.9\Gamma_0$. By applying a gradient field $6\Gamma_0/\lambda$, the emission photon spectra are shown in Fig. 5 where the red solid line is the reflection spectrum and the blue dotted line is the transmission spectrum. There are two reflection peaks which correspond to the two emitters. The separation between the two peaks is $12.6\Gamma_0$ from which we can determine that the emitter separation is $2.1\lambda$ which matches the given value very well. The FWHM linewidth of the left reflection peak is $1.45\Gamma_0$ and the maximum reflectivity is $0.45$. We therefore have $\Gamma_1+\gamma_1=1.45\Gamma_0$ and $(1+\gamma_1/\Gamma_1)^{-2}=0.45$ from which we can determine that $\Gamma_1=0.97\Gamma_0$ and $\gamma_1=0.48\Gamma_0$ which also match the given values $\Gamma_0$ and $0.5\Gamma_0$ respectively very well. Similarly, the FWHM linewidth of the right reflection peak is $2.46\Gamma_0$ and the maximum reflectivity is $0.41$. We have $\Gamma_2+\gamma_2=2.46\Gamma_0$ and $(1+\gamma_2/\Gamma_2)^{-2}=0.41$. We can then determine that $\Gamma_2=1.58\Gamma_0$ and $\gamma_2=0.88\Gamma_0$ which also match the given values very well.

\section{Summary}

In summary, we have proposed a method to determine the emitter separation in a waveguide-QED system even if the emitter separation is in the deep sub-wavelength scale. For a high quality photonic waveguide with negligible decay to the free space, the emitter separation can be deduced from the reflection dip position. For a waveguide with decay to the free space, the emitter separation can be determined from the dipole-dipole splitting. If there are more than two emitters which are very close to each other, we can also determine the number of emitters by simply counting the number of reflection dips or the reflection peaks. Moreover, we also show how to measure ultrasmall emitter separation change in the waveguide-QED system. This may then be used to measure the strain or temperature variation with both high sensitivity and spatial resolution. We also show how to measure the decay rates to the waveguide even if the emitters have different coupling constants. Our theory here may find important applications in designing the waveguide-QED-based device and sensor.

\section*{Acknowledgments}
We thank X. Zeng, M. T. Cheng, and C. P. Sun for helpful discussion.  This research is supported by NPRP Grant No. 8-352-1-074 from the Qatar National Research Fund (QNRF) and a grant from King Abdulaziz City for Science and Technology (KACST).

\end{document}